\documentclass{article}
\usepackage{amssymb}
\usepackage{amsmath}

\input{tcilatex}

\begin{document}

\begin{center}
{\LARGE Gravitational waves versus black holes}

Trevor W. Marshall, School of Mathematics, University of Manchester,
Manchester M13 9PL, UK

June 30, 2007
\end{center}

\QTP{Body Math}
\textbf{Abstract\qquad }It is argued that, in order for the gravitational
field to be propagated as a wave, it is necessary for it to satisfy a
further set of field equations, in addition to those of Einstein and
Hilbert, and these equations mean there is a preferred coordinate frame,
called the Global Inertial Frame, giving rise to a unique metric . The
implication is that a true gravitational field is not compatible with
Einstein's Principle of Equivalence, which is in contradiction with his
other fundamental concept of locality. The additional field equations ensure
that gravitational collapse does not go below the Schwarzschild radius,
thereby excluding the possibility of singular solutions (black holes) of the
Einstein-Hilbert equations. Such solutions would also violate Einstein's
locality principle.

\section{Introduction}

In 1907, in two articles\cite{firstloc}\cite{firstpe} separated by a few
months, Albert Einstein stated two principles to which he would subsequently
give the names Principle of Local Action (PLA), or in the original Prinzip
der Nahewirkung, and Principle of Equivalence (PE). The first of these
states that nothing goes faster than light, and it was considered by him to
be the principle underlying \ what we now call Special Relativity,
Einstein's theory of 1905. He restated it many times during his later life%
\cite{epr}\cite{dialect}\cite{autobio}, especially in his criticism of
quantum theory, and in the last of these citations he stated the view that
it would be impossible to do science without it. As for PE, although he
classified it as "the happiest thought of my life"(\cite{pais} p178),
leading him, as it did, to the 1915 theory which we now know as General
Relativity (GR), the statement of 1907 was very modest compared with what it
later became; it stated that no observer can distinguish between a uniformly
accelerated frame and a frame at rest in a uniform gravitational field. 

Even before the creation of GR, the two principles presented some
compatibility problems; both the gravitational red shift and the variation
of the free-space refractive index, predicted by Einstein in 1911 and
resulting in the bending of light beams, made the limiting value of $c$
somewhat ambiguous. But the search from then to 1915, by Einstein in
collaboration with Grossman\cite{pais} and independently by Hilbert\cite%
{mehra}, for a group of transformations more general than that of Lorentz
under which the laws of field transmission should be invariant, led
eventually to a very strong formulation of PE. From that point on Einstein
was to state repeatedly that all the laws of physics should be independent
of the coordinate system. This strong form of PE is, I shall argue, totally
destructive of PLA.

During the formative years of GR he and Grossman tried to create a theory of
gravitation based on a more restricted group of transformations\cite{pais}
than is demanded by PE. An examination of the articles they wrote during
this period shows that it was a desire to incorporate PLA, in the form of a
tensor representing the local energy density of the gravitational field,
that motivated the various twists and turns taken by them. Einstein had
decided by 1915 that this was not possible, and in articles by Hilbert\cite%
{hilbert} and Schr\"{o}dinger\cite{schrod} the impossibility of constructing
such a tensor was confirmed. Nevertheless, he returned in 1918 to the idea
that gravitation, like any other field, must propagate at the velocity $c$.
By imposing on the gravitational potential, which in conformity with PE is
the same as the metric of the curved space, a certain restriction, he
deduced a formula\cite{gravwave} for the \emph{total }radiation emitted even
though the \emph{local} flux thereof remained ambiguous. It was not long
before Eddington\cite{edding} drew attention to the fact that Einstein's
waves could be transformed out of existence by a coordinate change permitted
under PE; he observed caustically that the waves "travel at the speed of
thought". However, both Hilbert and Eddington (\cite{edding} pp40-41) showed
themselves ready to modify PE in order to accommodate some notion of
locality. The former\cite{mehra} actually identified the related notion of
"causality" as a further requirement for a physical theory of gravitation,
and he stated a criterion to meet this requirement, that the signature of
the metric $g_{ij}$ must be preserved as $\left( +---\right) $ throughout
any evolution of a system. I shall show that Hilbert's causality is a close
relation of Einstein's PLA, and that playing fast and loose with the
signature of $g_{ij}$ is precisely what has led a substantial part of our
community into the blind alley of black-hole theory. It should be noted that
more recently\cite{lm} it was shown that coordinate transformations exist
which transform away the Einstein energy loss globally as well as locally. A
strong PE is not compatible with the existence of gravitational waves.

With respect to black holes Oppenheimer and Snyder\cite{oppsny} showed in
1939 that a fairly simple metric describes what may plausibly be considered
a set of dust particles under zero pressure which, in a finite time
interval, collapses into a black hole, so called because light cannot escape
from it. Objects with nonzero pressure are now widely believed to undergo
similar collapse, subject to certain conditions of size and internal
temperature, but the dust cloud remains the basis of the black-hole
paradigm. It seems likely that Einstein knew of this work before it was
published, because almost simultaneously he published an article\cite{einvbh}
criticizing the concept, on the grounds that the crossing of the event
horizon by the dust particles would violate his locality principle. In this
he was also affirming his support for Hilbert's notion of "causality",
referred to above; at the point of crossing the event horizon, the signature
of the metric changes from $\left( +---\right) $ to $\left( -+--\right) $.
In other words PE, a principle which Einstein apparently still supported,
predicted a process which violated locality; it seems that locality won the
day as far as he was concerned! \bigskip 

\section{The relativistic theory of gravitation}

GR has only one set of field equations%
\begin{equation}
\mathcal{R}^{ij}-\frac{1}{2}g^{ij}g_{kl}\mathcal{R}^{kl}=8\pi \kappa
T^{ij}\quad ,\quad \kappa =\frac{G}{c^{2}}\quad ,  \label{einhil}
\end{equation}%
where $\mathcal{R}^{ij}$ is the contracted curvature tensor derived from the
metric tensor $g_{ij}$, and $T^{ij}$ is the local material stress, or energy
density \ tensor. In his article on gravitational waves\cite{gravwave}
Einstein introduced the noncovariant condition%
\begin{equation}
\partial _{i}\Phi ^{ij}=0\quad ,\quad \Phi ^{ij}=g^{ij}\sqrt{-g}\quad .
\label{gauge}
\end{equation}%
This is often referred to nowadays as a \emph{gauge condition}, and it was
further developed by\ de Donder\cite{deDond} and later by Fock\cite{fock}, \
who showed that, if the coordinates are cartesian and satisfy (\ref{gauge}),
then they also satisfy the \emph{harmonic condition}%
\begin{equation}
\square x_{i}=0\quad ,
\end{equation}%
where $\square $ is the generalized d'Alembertian operator%
\begin{equation}
\square =\frac{1}{\sqrt{-g}}\partial _{j}g^{jk}\sqrt{-g}\partial _{k}\quad .
\end{equation}%
Fock showed that many complex calculations in GR are greatly simplified in
harmonic coordinates. In particular he used them to calculate the rate at
which a binary system, of which the only example for which he then had the
relevant data was Sun-Jupiter, radiated away energy through gravitational
waves. Although the effect was too small to observe in that system, his
calculation formed the basis for such a system involving a neutron star,
which made possible the calculation\cite{binpuls} of the radiative energy
loss in a binary pulsar.

Although Fock criticized PE, and indeed considered that it was incorrect to
label the Einstein-Hilbert theory "General Relativity", he did not consider
the harmonic system of coordinates as in any way privileged. That was left
to a later school of gravity theorists in the Soviet Union\cite{lm} who
proposed that, within a given family of metrics classified as equivalent
under PE, it is the harmonic one, which correctly describes the physical
system. There is still a family of coordinate frames with this metric, but \
the group of coordinate transformations connecting these frames is the
Lorentz group of the Special Theory of Relativity. They called the resulting
theoretical structure the \emph{\ Relativistic Theory of Gravitation }(RTG).
A gravitating system studied in isolation, for which the distant field is
Newtonian, is classified as an \emph{island system}, and I shall call the
harmonic frame for such a system, in which its centre of mass is at rest,
the \emph{global inertial frame }(GIF)\emph{.}

RTG differs significantly from General Relativity. In particular
gravitational collapse does not go below the Schwarzschild radius, and so
there are no black-hole singularities. At the same time the theory includes
unambiguous expressions for the material and gravitational stress tensors,
and the latter is associated with a real physical field which propagates at
the speed of light, just like the electromagnetic field of Faraday and
Maxwell. The quantities $\Phi ^{ij}$ are considered to be gravitational 
\emph{fields }instead of being the gravitational \emph{potentials }of GR,
and the field equations of RTG consist of (\ref{einhil}) \emph{together with 
}(\ref{gauge}). \ It is a radical departure to treat this latter equation as
an essential, globally valid field equation, which, for a given source,
results in a unique field, defined with respect to the GIF.

A simple example of an island frame arises from considering the
Schwarzschild metric for a point mass $m$ at the origin%
\begin{equation}
ds^{2}=\frac{r-2m}{r}dt^{2}-\frac{r}{r-2m}dr^{2}-r^{2}\left( d\theta
^{2}+\sin ^{2}\theta d\phi ^{2}\right) \quad ,
\end{equation}%
for which the d'Alembertian is%
\begin{equation}
\square =\frac{r}{r-2m}\partial _{t}^{2}-\frac{1}{r^{2}}\left[ r\left(
r-2m\right) \partial _{r}^{2}+2\left( r-m\right) \partial _{r}+\partial
_{\theta }^{2}+\cot \theta \partial _{\theta }+\csc ^{2}\theta \partial
_{\phi }^{2}\right] \quad .
\end{equation}%
The coordinate $z=r\cos \theta $ is not harmonic, while $Z=\left( r-m\right)
\cos \theta $ is, so the metric associated with the GIF is%
\begin{equation}
ds^{2}=\frac{R-m}{R+m}dt^{2}-\frac{R+m}{R-m}dR^{2}-\left( R+m\right)
^{2}\left( d\theta ^{2}+\sin ^{2}\theta d\phi ^{2}\right) \quad .
\label{schwharm}
\end{equation}%
Using this metric instead of Schwarzschild's gives a small correction\cite%
{lm} to the perihelion advance of a planet's elliptic orbit, to the bending
of light, and to the gravitational red shift, but they are all far too small
to be measured.

\section{The gravitational collapse of a dust cloud}

Oppenheimer and Snyder\cite{oppsny} studied the metric%
\begin{equation}
ds^{2}=d\tau ^{2}-S_{R}^{2}dR^{2}-S^{2}\left( d\theta ^{2}+\sin ^{2}\theta
d\phi ^{2}\right) \quad ,
\end{equation}%
where%
\begin{equation}
S=\left[ R^{3/2}-\frac{3}{2}\tau f\left( R\right) \right] ^{2/3}\quad ,
\end{equation}%
and 
\begin{equation}
f\left( R\right) =\left\{ 
\begin{array}{c}
1\quad \left( R>1\right) \quad , \\ 
R^{3/2}\quad \left( R<1\right) \quad .%
\end{array}%
\right. 
\end{equation}%
Note that%
\begin{equation}
S_{R}=\left\{ 
\begin{array}{c}
\sqrt{R/S}\quad \left( R>1\right) \quad , \\ 
S/R\quad \left( R<1\right) \quad ,%
\end{array}%
\right. 
\end{equation}%
which means that this metric has a discontinuity at $R=1$. They showed that
a frame defined by it is comoving in the sense that its material stress
tensor,%
\begin{equation}
T^{ij}=\frac{1}{8\pi }\left( \mathcal{R}^{ij}-\frac{1}{2}g^{ij}g_{kl}%
\mathcal{R}^{kl}\right) \quad ,
\end{equation}%
has only a timelike nonzero component, that is 
\begin{equation}
T^{\alpha \beta }=T^{\alpha 0}=0\quad ,\quad T^{00}=\frac{ff^{\prime }}{2\pi
S^{2}S_{R}}\quad ,
\end{equation}%
for which the mass density is%
\begin{equation}
\rho dRd\theta d\phi =T^{00}\sqrt{-g}dRd\theta d\phi =\frac{ff^{\prime }}{%
2\pi }\sin \theta dRd\theta d\phi \quad .
\end{equation}%
They used this metric to model the gravitational field of a spherically
symmetric dust cloud with no rotation; the function $f\left( R\right) $
contains the $R$-dependence of $\rho $, and in the exterior region, $R>1$,
it is constant, making $\rho $\ zero. The comoving property of this frame
means that the timelike coordinate $\tau $ is a local proper time. It is a
simple matter to establish that each particle in the cloud reaches the
"event horizon", a point at which the $\left( +---\right) $ signature of
space-time changes, within a finite interval of $\tau .$ This was taken in 
\cite{oppsny} to show that the cloud collapses into a black hole, but we
shall show, using the harmonic coordinates, that this was an incorrect
conclusion. 

For the exterior region, $R>1$, the latter authors looked for a coordinate
transformation $\left( \tau ,R\right) \rightarrow \left( t,r\right) $ such
that the above metric transforms into the Schwarzschild metric with $2m=1$.
This leads to the result 
\begin{equation}
t=\frac{2}{3}\left( R^{3/2}-S^{3/2}\right) +2-2\sqrt{S}+2\log \frac{\sqrt{S}%
+1}{2}-\log \left( S-1\right) \quad ,\quad S=r\quad .  \label{extsol}
\end{equation}%
They then obtained an interior metric by imposing the conditions $g^{tr}=0$
and $S=r$ there also. They required in addition that $t$, but not $t_{R}$,
be continuous at $R=1$, and they deduced that%
\begin{equation}
t=-\log \left( y-1\right) +2\log \frac{\sqrt{y}+1}{2}+\frac{8}{3}-\frac{2}{3}%
y^{3/2}-2\sqrt{y}\quad ,  \label{intsol}
\end{equation}%
where%
\begin{equation}
y=\frac{R^{2}-1}{2}+\frac{r}{R}\quad .
\end{equation}%
The coordinates $\left( t,r\right) $ they refer to as "external coordinates"
for both the exterior and the interior regions; this is by contrast with the
comoving coordinates $\left( \tau ,R\right) $. In the external frame $t$
becomes infinite for the exterior at $r=r_{\infty }=1$ and for the interior
at%
\begin{equation}
r_{\infty }=\frac{3}{2}R-\frac{1}{2}R^{3}\quad \left( R<1\right) \quad ,
\end{equation}%
which is the event horizon\emph{.} 

All the particles of the cloud go to their final positions, given by $%
r_{\infty }\left( R\right) $ and stay there$;$\emph{\ this is a stationary
state and it is surprising that Oppenheimer and Sneider did not acknowledge
it as such}. What happens on the wrong side of the event horizon, that is
for proper times satisfying%
\begin{equation}
\tau _{\infty }>\frac{2}{3}-\frac{2}{3}\left( \frac{3-R^{2}}{2}\right)
^{3/2}\quad ,
\end{equation}%
has no physical relevance at all. What seems to have caused their confusion
was the discovery that the metric coefficients become singular near $r=1.$
There is a simple explanation for this, since the Schwarzschild metric in
the exterior region has the same behaviour. Furthermore, the density $\rho
\left( r_{\infty }\right) $ becomes infinite at $r_{\infty }=1,$ as may be
shown from the constant cumulative density in the comoving frame, namely%
\begin{equation}
P\left( R\right) =\int_{0}^{R}4\pi \rho \left( R\right) dR=R^{3}\quad .
\end{equation}%
The corresponding quantity in the external frame is%
\begin{equation}
P\left( r_{\infty }\right) =\left[ R\left( r_{\infty }\right) \right]
^{3}=8\cos ^{3}\left[ \frac{\pi }{3}+\frac{\arccos r_{\infty }}{3}\right]
\quad ,
\end{equation}%
and near $r_{\infty }=1$ this gives an infinite density, namely%
\begin{equation}
4\pi \rho _{\infty }\left( r\right) \sim \sqrt{\frac{3}{2-2r}}\quad .
\end{equation}%
The infinite value of $g_{rr}$ found by Oppenheimer and Sneider in the limit 
$r\rightarrow 1$ is simply a consequence of this mass concentration at the
surface of the cloud. By contrast $g_{tt}$ is zero at all points on the
event horizon, which indicates an infinite red shift in the limit $%
t\rightarrow \infty $.

Their exterior solution (\ref{extsol}) is not unique, owing to the arbitrary
function in their eqn.19 having been equated to one. Of course the absence
of uniqueness is inevitable in General Relativity, it being a manifestation
of PE, which ordains that no coordinate system is to be favoured. In RTG, by
contrast, the external solution is unique, because it has to satisfy the
harmonic condition; in that case the external solution is identical with (%
\ref{extsol}), except (see the previous section) that $S$ is equal to $r+1/2$
instead of to $r$. As for the internal solution, the condition $g^{tr}=0$ of
Oppenheimer and Snyder somewhat fortuitously ensures that $t_{R}/r_{R}$ is
continuous at $R=1$ if $r$ and $t$ are, and this then makes $g_{tt}$ and $%
g_{rr}$ continuous there, even though $g_{RR}$ is not. The condition $%
g^{tr}=0$ is arbitrary though allowable under PE, but it is not satisfied by
the RTG solution. That is constructed by imposing the harmonic condition,
and by requiring the continuity of both $r_{R}$ and $t_{R}$ at $R=1$,
thereby setting a constraint on the function $f(R)$ in the internal region;
this ensures by a different route the continuity of the metric.

In accordance with the latter condition, I choose for the interior density
function%
\begin{equation}
f\left( R\right) =R^{3/2}e^{3X/2}\quad ,\quad X=1-R\quad \left( R<1\right) 
\end{equation}%
giving%
\begin{equation}
S_{R}=S\xi \quad ,\quad \xi =\frac{X}{R}+\sqrt{\frac{R^{3}}{S^{3}}}\quad
\left( R<1\right) \quad ,
\end{equation}
and the mass density%
\begin{equation}
\rho dRd\theta d\phi =\frac{3XR^{2}e^{3X}}{8\pi }\sin \theta dRd\theta d\phi
\quad .
\end{equation}%
The d'Alembertian operator in terms of the coordinates $\left( R,S,\theta
,\phi \right) $ is%
\begin{equation}
\square =\square _{1}-\frac{1}{S^{2}}\left( \partial _{\theta }^{2}+\cot
\theta \partial _{\theta }+\csc ^{2}\theta \partial _{\phi }^{2}\right)
\quad ,
\end{equation}%
where

\begin{eqnarray}
\square _{1} &=&\left( \frac{f^{2}}{S}-1\right) \left( \partial _{S}^{2}+%
\frac{1}{S}\partial _{S}\right) -\frac{2}{\xi S}\partial _{R}\partial _{S}-%
\frac{1}{\xi ^{2}S^{2}}\partial _{R}^{2}+\left[ \frac{ff^{\prime }}{\xi S^{2}%
}-\frac{1}{S}\right] \partial _{S}  \notag \\
&&-\frac{1}{\xi ^{2}S^{2}}\left[ \frac{5\xi }{2}-\frac{f^{\prime }}{f}-\frac{%
\xi _{R}}{\xi }\right] \partial _{R}\quad .
\end{eqnarray}

The harmonic coordinates%
\begin{equation*}
x_{i}=\left( t,r\sin \theta \cos \phi ,r\sin \theta \sin \phi ,r\cos \theta
\right) 
\end{equation*}%
satisfy 
\begin{equation}
\square _{1}t=0\quad ,  \label{tharm1}
\end{equation}%
and%
\begin{equation}
\left( \square _{1}+\frac{2}{S^{2}}\right) r=0\quad ,  \label{rharm1}
\end{equation}%
for which the exterior solutions, obtained by putting $f=1$ and $\xi =\sqrt{%
R/S^{3}}$, are the same as (\ref{extsol}) for $t$, with $r$ replaced by%
\begin{equation}
r=S-\frac{1}{2}\quad .
\end{equation}%
This is, as expected, the harmonic form of the Schwarzschild solution, as
obtained already in the previous section. The interior coordinates are
obtained by solving these same equations with $r,t,r_{R}$ and $t_{R}$
continuous at $R=1$, and $t$ and $r$ having the Correspondence Principle
(CP) behaviour $t\sim \tau ,r\sim S$ as $S$ tends to $+\infty $. The latter
conditions give the leading term for large positive $S,$ namely%
\begin{equation}
t\sim -\frac{2}{3}S^{3/2}e^{3Z/2}\sim -\frac{2}{3}r^{3/2}e^{3Z/2}\quad
,\quad e^{-Z}=e^{X}R\quad ,
\end{equation}%
or equivalently%
\begin{equation}
r\sim r_{0}e^{-Z}\quad ,\quad r_{0}=\left( -\frac{3t}{2}\right) ^{2/3}\quad ,
\end{equation}%
leading to a density%
\begin{equation}
\rho \left( r\right) \sin \theta drd\theta d\phi \sim \frac{3r^{2}}{4\pi
r_{0}^{3}}\sin \theta drd\theta d\phi \quad \left( r<r_{0}\right) \quad .
\end{equation}%
This gives us the collapse of a sphere of radius $r_{0}\left( t\right) $
with initially uniform mass density, and we see that the uniform density is
preserved during the first stage of collapse, in agreement with the
Newtonian behaviour for such a density. The leading relativistic correction
is obtained from the next term, namely 
\begin{equation}
t\sim -\frac{2}{3}S^{3/2}e^{3Z/2}-S^{1/2}\left( \frac{3}{2}e^{Z/2}+\frac{1}{2%
}e^{-3Z/2}\right) \quad ,\quad r\sim S-\frac{3}{4}e^{-Z}+\frac{1}{4}e^{-3Z}
\end{equation}%
from which $S$ may be eliminated to give%
\begin{equation}
r\sim e^{-Z}\left( r_{0}+\frac{1}{4}-\frac{1}{4}e^{-2Z}\right) \quad ,\quad
r_{0}=\left( -\frac{3t}{2}\right) ^{2/3}-\frac{5}{2}\quad ,
\end{equation}%
with the density%
\begin{equation}
\rho \left( r\right) \sin \theta drd\theta d\phi \sim \frac{3r^{2}}{4\pi
r_{0}^{3}}\left( 1-\frac{3}{4r_{0}}+\frac{5r^{2}}{4r_{0}^{3}}\right) \sin
\theta drd\theta d\phi \quad \left( r<r_{0}\right) \quad .  \label{reldens}
\end{equation}%
This latter result shows a departure from the uniform density associated
with Newtonian gravity; as the collapse progresses the density near the
surface $\left( r=r_{0}\right) $ increases faster than at the centre of the
cloud.

To follow the evolution into the region of strong gravity requires a
numerical approach, for which we make the further change of variables from $R
$ to $X$ and $S$ to $Y=\log S+Z$, leading to 
\begin{align}
-S^{2}\xi ^{2}\square _{1}& =\partial _{X}^{2}-2e^{-3X/2-3Y/2}\partial
_{X}\partial _{Y}+\left( e^{-3X-3Y}-\xi ^{2}R^{2}e^{2X-Y}\right) \partial
_{Y}^{2}  \notag \\
& +\left( \xi ^{2}+\frac{1}{R^{2}}-\frac{3\xi XR}{2}e^{2X-Y}\right) \partial
_{Y}  \notag \\
& -\left[ \frac{5\xi }{2}+\frac{3}{2}-\frac{3}{R}+\frac{2+3X}{2\xi R^{2}}%
\right] \left( \partial _{X}+\frac{X}{R}\partial _{Y}\right) 
\end{align}%
From this it is possible to derive more terms in the asymptotic series for $r
$ and $t$, namely%
\begin{equation}
t\sim -\frac{2}{3}e^{3Y/2}-\left( \frac{3}{2}+\frac{1}{2}e^{-2Z}\right)
e^{Y/2}+\sum_{0}^{\infty }t_{n}e^{-nY/2}\quad ,
\end{equation}%
and%
\begin{equation}
r\sim e^{-Z}\left( e^{Y}+\sum_{0}^{\infty }u_{n}e^{-nY/2}\right) \quad ,
\end{equation}%
where the first few nonzero terms are

\begin{eqnarray}
t_{0} &=&\frac{2}{3}e^{-3X/2}\quad ,  \notag \\
t_{1} &=&\frac{2}{5}e^{Z}+\frac{15}{8}-\frac{1}{4}e^{-2Z}-\frac{1}{40}%
e^{-4Z}\quad ,  \notag \\
t_{2} &=&\frac{1}{2}e^{-3X/2}+\frac{3}{2}e^{X/2}\left( 13-6X+X^{2}\right)
-20\quad ,  \notag \\
t_{3} &=&\frac{43}{70}e^{Z}+\frac{29}{96}-\frac{1}{10}e^{-Z}-\frac{5}{32}%
e^{-2Z}+\frac{1}{160}e^{-4Z}+\frac{1}{3360}e^{-6Z}\quad ,
\end{eqnarray}%
and%
\begin{eqnarray}
u_{0} &=&-\frac{3}{4}+\frac{1}{4}e^{-2Z}  \notag \\
u_{3} &=&20-\frac{1}{2}e^{-3X/2}-\frac{3}{2}e^{X/2}\left( 13-6X+X^{2}\right)
=-t_{2}\quad ,  \notag \\
u_{6} &=&\frac{5}{6}-\frac{15}{4}\left( X^{2}-4X+9\right) e^{-X}-\frac{5}{12}%
e^{-3X}+\frac{100}{3}e^{-3X/2}\quad .
\end{eqnarray}

\bigskip The partial differential equations (\ref{tharm1}) and (\ref{rharm1}%
) must now be integrated subject to the boundary conditions at $X=0$, for
which the first equation displays a logarithmic singularity at $S=1$, that
is $Y=0$. This suggests there is a similar singularity along the
characteristic $Y=Y_{0}\left( X\right) $ through $\left( 0,0\right) $. This
is in fact the event horizon in the $\left( X,Y\right) $ coordinates, and it
is determined by the ordinary differential equation%
\begin{eqnarray}
Y_{0}^{\prime } &=&R\xi e^{X-Y_{0}/2}-e^{-3X/2-3Y_{0}/2}  \notag \\
&=&Xe^{X-Y_{0}/2}+\left( 1-X\right) e^{-X/2-2Y_{0}}-e^{-3X/2-3Y_{0}/2}\quad ,
\label{rtgchar}
\end{eqnarray}%
which integrates numerically to give the function

\begin{tabular}{|l|l|l|l|l|l|l|l|l|l|l|}
\hline
$X$ & 0.1 & 0.2 & 0.3 & 0.4 & 0.5 & 0.6 & 0.7 & 0.8 & 0.9 & 1.0 \\ \hline
$Y_{0}$ & .005 & .021 & .049 & .091 & .150 & .227 & .326 & .447 & .591 & .758
\\ \hline
\end{tabular}

It may be noted that this is finite at $X=1$, even though some coefficients
of the partial differential equations are infinite there. This is a
consequence of the choice made for the variable $Y$. A further advantage of
this choice is that, as $Y\rightarrow +\infty $, the characteristics are $Y=$%
constant, which aids the numerical integration. The latter is achieved by
replacing each partial differential equation by a set of ordinary
differential equations for the vectors $\mathbf{r}\left( X\right) =r\left(
X,Y_{i}\right) $ and $\mathbf{t}\left( X\right) =t\left( X,Y_{i}\right) ,$
where $Y_{i}$ are a set of 100 values of $Y$ spaced equally between an upper
value $Y_{0}$ at which the above asymptotic expansions apply (I took $%
Y_{0}=5)$ and a lower value $Y_{1}$ close to zero (I took $Y_{1}=$ .001$)$.
\ The integration confirms that $t$ becomes infinite as $Y$ approaches $%
Y_{0}\left( X\right) $, which means that we are seeing an extreme form of
the gravitational red shift.\ The values of $r\left( X\right) $ along this
same curve, which we designate $r_{\infty }\left( X\right) $, are

\begin{tabular}{|l|l|l|l|l|l|l|l|l|l|l|l|}
\hline
$X$ & 0 & 0.1 & 0.2 & 0.3 & 0.4 & 0.5 & 0.6 & 0.7 & 0.8 & 0.9 & 1.0 \\ \hline
$r_{\infty }$ & .500 & .500 & .500 & .499 & .497 & .492 & .481 & .456 & .398
& .272 & 0 \\ \hline
\end{tabular}

and this shows that gravitational collapse ceases at the radius $r_{\infty 0}
$, which takes the value $0.5$ in our units$.$ The latter function, or
rather its inverse $X\left( r_{\infty }\right) $, gives us the limiting
density $\rho _{\infty }\left( r\right) $, since the density in terms of the
comoving coordinate $X$ remains constant throughout, so that the cumulative
density is%
\begin{equation}
P_{\infty }\left( r\right) =\int_{0}^{r}4\pi \rho _{\infty }\left( r^{\prime
}\right) dr^{\prime }=\left[ 1-X\left( r_{\infty }\right) \right]
^{3}e^{3X\left( r_{\infty }\right) }\quad .
\end{equation}%
A corresponding uniform density would give%
\begin{equation}
P_{U}\left( r\right) =\frac{r_{\infty }^{3}}{r_{\infty 0}^{3}}=8r_{\infty
}^{3}\quad ,
\end{equation}%
and a comparison may be made by plotting $P_{\infty }/P_{U}$ as a function
of $r$. This shows that the process indicated by (\ref{reldens}) above,
namely a concentration of particles at the surface, continues and
intensifies. For example, we find that $P_{U}$ would give less than 2\% of
the particle distribution in the range $.497<r<.5$, as compared with $%
P_{\infty }$ giving more than 28\%, while, in $.481<r<.5$, $P_{U}$ gives
less than 11\%, compared with $P_{\infty }$ giving more than 61\%. As $%
r_{\infty }$ tends to 0.5 the derivative of $P_{\infty }$, that is $\rho
_{\infty }$ becomes infinite; this is natural, because the metric
coefficient $g_{rr}$ must become infinite at $r=0.5$ to match with its value
in the exterior region $R>1$. Also, associated with the infinite red shift
at the event horizon, the determinant $g$ of the metric approaches zero
there.

\section{\protect\bigskip Conclusion}

Our result shows that, contrary to the general result claimed by \cite%
{oppsny}, there is, for a suitable choice of the density function $f\left(
R\right) ,$ and of course a suitable coordinate frame, a stationary solution
of the Einstein-Hilbert equation, and this state is approached in the limit $%
t\rightarrow +\infty $. It should be noted that, both in the present article
and in \cite{oppsny}, an additional condition had to be imposed in order to
obtain an interior metric satisfying the Correspondence Principle, and that
in both cases there is a singularity in the metric in the limit $%
t\rightarrow +\infty $.

More significant, however, is that the insistence on a global inertial frame
as the vehicle for gravitational waves restores the gravitational theory of
Einstein to the framework of his Special Theory. In particular the waves
travel with the velocity of light and are consistent with his principle of
locality. It is entirely natural then that RTG should also rule out
gravitational collapse beyond the event horizon, leading, as it would, to an
exotic topology in which the space and time coordinates would change places
and to a complete demolition of causality.

The key to understanding what happens as a particle approaches the event
horizon is that such an intense gravitational field produces a red shift
which becomes infinite, that is \emph{all physical processes including
gravitational collapse are infinitely slowed down}. Effectively time is
frozen. There is also an infinite mass density at the surface of the cloud.
This is an extreme situation which none of us is likely to experience
directly, but it ensures that, even in such circumstances, the world remains
local and comprehensible.


\begin{thebibliography}{99}
\bibitem{firstloc} A. Einstein, \emph{Ann. der Physik, }\textbf{23, }371-384
(1907)

\bibitem{firstpe} A. Einstein, \emph{Jahrb. der Radioacktiv. und Elektronik, 
}\textbf{4, }411-462 (1907)

\bibitem{epr} A. Einstein, B. Podolsky and N. Rosen, \emph{Phys. Rev., }%
\textbf{47}, 777 (1935)

\bibitem{dialect} A. Einstein, \emph{Dialectica, }\textbf{2, }320 (1948).
English  translation in \emph{The Born Einstein Letters, }Macmillan, London
(1971)

\bibitem{autobio} A. Einstein, in \emph{Albert Einstein:
Philosopher-Scientist, }ed. P. A. Schilpp, Tudor New York\emph{\ }(1949)

\bibitem{pais} A. Pais, \emph{Subtle is the Lord\ldots the Science and Life
of Albert Einstein, }University Press, Oxford (1982)\emph{\ }

\bibitem{mehra} J. Mehra, \emph{Einstein, Hilbert and the Theory of
Gravitation, }D. Reidel, Boston (1974)

\bibitem{hilbert} D. Hilbert, \emph{Goett. Nachrichten }\textbf{4}, 21 (1917)

\bibitem{schrod} E. Schr\"{o}dinger, \emph{Phys. Z., }\textbf{19, }4 (1918)

\bibitem{gravwave} A. Einstein, \emph{Sitzungsber. preuss. Akad. Wiss.,} 
\textbf{1}, 154 (1918)

\bibitem{edding} A. S. Eddington, \emph{The Mathematical Theory of
Relativity, }University Press, Cambridge (1924)

\bibitem{lm} A. Logunov and M \ Mestvirishvili, \emph{The Relativistic
Theory of Gravitation}, Mir, Moscow (1989)

\bibitem{oppsny} J. R. Oppenheimer and H. Snyder, \emph{Phys. Rev.\ }\textbf{%
56}, 455 (1939)

\bibitem{einvbh} A. Einstein \emph{Ann. Math. }\textbf{40}, 922 (1939)

\bibitem{deDond} T. de Donder, \emph{Th\'{e}orie des Champs Gravitiques, }%
Gauthiers-Villars, Paris\emph{\ }(1926)

\bibitem{fock} V. A. Fock, \emph{The Theory of Space, Time and Gravitation, }%
Pergamon, New York (1959)

\bibitem{binpuls} M. Walker and C. M. Will, \emph{Phys. Rev. Lett., }\textbf{%
22, }1741 (1980)
\end{thebibliography}
\end{document}